# DYNAMICAL THEORY OF GROUPS AND CLUSTERS OF GALAXIES

Gary A. Mamon

DAEC, Observatoire de Paris-Meudon, F 92195 Meudon, FRANCE

astro-ph/9308032   24 Aug 93

# DYNAMICAL THEORY OF
# GROUPS AND CLUSTERS OF GALAXIES

Gary A. Mamon

DAEC, Observatoire de Paris-Meudon, F 92195 Meudon, FRANCE

**Abstract.** The different dynamical processes (relaxation, dynamical friction, tides and mergers) operating in groups and clusters are reviewed. The small-scale substructure observed in clusters is argued to be the remnants of the cores of rich clusters that merged together, rather than large groups falling into the cluster. The *ROSAT* X-ray observations of two groups of galaxies are discussed, and, contrary to a previous claim, the baryon fraction is high, relative to the constraints from baryonic nucleosynthesis. A general theory of the fundamental surface of groups is presented, allowing one to determine with reasonable confidence the precise cosmo-dynamical state of a given group of galaxies. The data from groups is then consistent with a universal true $M/L$ of $440\,h$, roughly 4 times larger than previous estimates, the discrepancy occurring because most groups are still relatively near cosmological turnaround. This high $M/L$ and the young cosmo-dynamical state of groups suggests a density parameter $\Omega > 0.3$. Hickson's compact groups are explained as a mixture of virialized groups, loose groups near full collapse, and chance alignments from collapsing loose groups. Finally, the level of projection effects contaminating samples of binary galaxies within groups is shown to be important.

## 1. Introduction

Thanks to gravity, galaxies like to congregate in groups and clusters. As seen in Table 1 below, only a minority of galaxies seem to live in isolation. From the general hierarchical clustering of galaxies in the Universe, one can separate the various systems of galaxies, according to richness (number of galaxies within given magnitude interval and distance from the system's center), *i.e.*, groups vs. clusters, compactness (mean surface brightness), *i.e.*, compact vs. loose groups, with an isolation criterion (compact groups and binaries). Note that with Abell's definition, one has $N_{\rm gal} = 4$, 44, and 106 for the Local Group (Milky Way, M31, M33 and the LMC), Virgo and Coma clusters, respectively.

Despite their relatively rare occurrence in the Universe, there has been plenty of studies of clusters of galaxies. Indeed, clusters are popular because they are the largest objects whose cores are in dynamical equilibrium, hence *virialized* (they obey the virial theorem), as contamination by interlopers is not too significant. In contrast, the outer regions of clusters are thought to still be feeling the effects of their *infall* onto the virialized cores, and to complicate matters even further, clusters often display substructure, as is well shown in pictures of the hot gas traced by the *ROSAT* satellite (*e.g.*, White, Briel & Henry 1993). This substructure is a tracer of cosmological parameters such as the density parameter, $\Omega$, and the spectrum of primordial density fluctuations (see §3).

## Table 1: 2D definitions of structures

|  | Criteria | $N_{\rm gal}$ | $\langle M/L \rangle$ | $f_{\rm gal}$ |
|---|---|---|---|---|
| Clusters | $m < m_3 + 2$<br>$hR < 1.5\,{\rm Mpc}$ | $30 - 300$ | $300\,h$ | $10\%$ |
| Loose Groups | $m < m_3 + 2$<br>$hR < 1.5\,{\rm Mpc}$ | $3 - 30$ | $150\,h$ | $50\%$ |
| Compact Groups | $m < m_1 + 3$<br>$\theta_n > 3\theta$ | $4 - 7$ | $50\,h$ | $0.1\%$ |
| Binaries | $m < m_1 + 3$<br>$\theta_n > 5\theta$ | $2$ | $100\,h$<br>$(100\,{\rm kpc})$ | $10\%$ |
| Isolated |  | $1$ |  | $30\%$ |

NOTES: The criteria are taken from Abell (1958), Hickson (1982), Turner (1976a), for the clusters, compact groups and binaries, respectively. Here, $N_{\rm gal}$ is the number of galaxies per system, $f_{\rm gal}$ is the fraction of galaxies in the Universe that belong to the type of system under consideration, and $h = H_0/(100\,{\rm km\,s^{-1}\,Mpc^{-1}})$.

Also, as dense systems near equilibrium, clusters represent an excellent laboratory to study dynamical interactions between galaxies, with the caveat that since their potential wells are deep, the relative encounter velocities are large, hence the interactions are short and not very damaging to the galaxies. One would then like to understand the segregation in morphologies, with elliptical galaxies predominantly occurring in the dense regions such as the cluster cores (see Mamon, in these proceedings), and the recent inference of high central concentration of dark matter relative to gas in clusters.

Loose groups have the advantage of being numerous, and for this reason, are often used as distance indicators, since if one knows that the distance to one object is known to some accuracy $\Delta D$, the distance obtained from $N$ galaxies believed to be all lying in the same group ought to be $\Delta D/\sqrt{N}$. Also, although not as extreme as clusters, loose groups can be thought to be good tracers of the Universe, and for many years, astronomers have tried to link the group mass-to-light ratios to $\Omega$ by simple extrapolation: $\Omega \simeq (M/L)/(1500h)$. And the distribution of their properties is again related to both $\Omega$ and the primordial density fluctuation spectrum, see §§4 and 5.

There are three serious problems with loose groups: 1) They suffer from important contamination from interlopers. 2) They are rarely virialized at best, so that the true mass-to-light ratio is a function of both the mass-to-light ratio obtained by assuming virial equilibrium and the cosmo-dynamical state of the group (expanding, collapsing, collapsed, virialized ...), see §5. 3) Groups could be biased $M/L$ tracers, if significant amounts of dark matter bound to the group lurk beyond the galaxies.

Compact groups appear so dense in projection that they would be the highest density isolated systems of galaxies, denser than the cores of rich clusters. Unfortunately they are very rare (see Table 1), and they may suffer from serious contamination from a surrounding loose group. This last point is a matter of debate (§6). If this contamination is low, then compact groups would indeed be extremely dense, and as such would serve as the ideal sites for strong galaxy evolution, by dynamical interaction, and by the star formation which this interaction may trigger. They would then also be extreme cosmological tracers, a little bit like the clusters, and thus allow one to determine $\Omega$ and the primordial density fluctuation spectrum. If, on the other hand, contamination by surrounding loose groups were indeed important, one still expects that the resulting chance alignments within loose groups will be binary-rich (Mamon 1992b), thus making compact groups interesting sites for galaxy evolution.

One could go on and state that binary galaxies are potential sites of strong galaxy evolution. Moreover, they are used to trace the matter distribution in galactic halos of dark matter, in particular the extent of these halos. The difficulty here is that again, contamination by chance alignments of galaxies within surrounding loose groups may be very important (see §7). Finally, one should state that isolated galaxies are very interesting as they serve as reference galaxies to which to compare the galaxies in denser environments.

## 2. Dynamical Processes

The reader is encouraged to read the excellent reviews on the details on the different dynamical processes by White (1983) and Richstone (1990), and the classic books by Saslaw (1985) and Binney & Tremaine (1987).

The dynamics of groups and clusters are set by their cosmological initial conditions. An homogeneous isolated system will first expand with the local Hubble flow. Then its high density will force it to decouple from the Hubble flow and it will reach its maximum expansion *turnaround*, collapse, and subsequently virialize. This equilibrium does not last forever, as virialization is followed by dissipation of orbital energy, caused by dynamical friction against an intergalactic background, and by tidal friction during collisions and merging. An inhomogeneous system will evolve in the same way, except that the denser regions will collapse and virialize first, and the low-density regions will later collapse onto the virialized core of the system (*secondary infall*) and subsequently virialize at a larger radius. Conservation of energy then yields a relation between the epoch of turnaround and the crossing time in virial equilibrium (Gunn & Gott 1972): $T_{\rm ta} = \pi t_{\rm cr}$ where the crossing time is defined as $t_{\rm cr} = (3/5)^{3/2} R_V/V_V$, where $R_V$ and $V_V$ are the virial radius and velocity dispersion, respectively.

This can be adapted to the circular orbital time:

$$\tau_{\rm circ} = \frac{2\pi R}{V_{\rm circ}(R)} = \left[\frac{3\pi}{G\bar{\rho}(R)}\right]^{1/2} .$$

As a test particle undergoes scattering collisions within a sea of field particles, it will progressively forget its initial conditions. This *two-body relaxation* time can be defined in at least three ways:

$$\tau_{2-{\rm rel}} \equiv \left\langle \frac{1}{v^2}\frac{dV^2}{dt} \right\rangle^{-1} \text{ or } \left\langle \frac{1}{E}\frac{dE}{dt} \right\rangle^{-1} \text{ or } \left\langle \frac{d\sin^2 \Delta\alpha}{dt} \right\rangle^{-1} ,$$

where $\Delta\alpha$ is the deflection angle in an encounter. Chandrasekhar (1942) has shown that this can be written as

$$\tau_{2-\mathrm{rel}} = \frac{v^3}{G^2 m_f^2 n f(v/\sigma_v) \ln \Lambda} \ ,$$

where $v$ is the velocity of the test particle, $m_f$, $n$, and $\sigma_v$ are the mass, number density, and 1D velocity dispersion of the field particles, respectively, $f$ is a function of order unity, and $\ln \Lambda$, also of order unity is called the Coulomb logarithm, where $\Lambda$ is the ratio of maximum to minimum impact parameter. For a system of galaxies and dark matter particles, one finds that the galaxies relax by galaxy-galaxy collisions, but not by collisions with individual dark matter particles (whose masses are too low). Similarly, the dark matter particles relax mainly by collisions with individual galaxies.

Gurzadyan & Savvidy (1984, 1986) estimated the *collective* relaxation time, obtained not by summing up the encounters but by computing the collective response of the system. They obtain

$$\tau_{N-\mathrm{rel}} = \mathrm{Cst} \, \frac{v}{G m_f n^{2/3}} \ .$$

This collective relaxation turns out to be somewhat more efficient than two-body relaxation in clusters and loose groups but not in dense groups. In general, only the cores of rich cluster are relaxed.

Lynden-Bell (1967) has shown that particles can rapidly forget their initial conditions if they evolve in a rapidly time varying potential:

$$\tau_{\mathrm{v-rel}} \sim \tau_{\mathrm{ff}} \sim \tau_{\mathrm{dyn}} \quad \mathrm{when} \quad |\frac{\partial \phi}{\partial t}| > |\mathbf{v} \cdot \nabla \phi| \ ,$$

where $\tau_{\mathrm{ff}}$ is the free-fall time, and $\phi$ is the global potential. This applies for example to collapsing systems, as is often the case in cosmology, and thus explains why the cores of elliptical galaxies appear relaxed although their 2-body (and collective) relaxation times are much longer than the age of the Universe.

Chandrasekhar (1943) also considered the effects of many scattering encounters on the forward velocity of a test particle. Because field particles are scattered in such a way that in the frame of the test particle, the field particle density is higher behind the test particle than in front of it. This leads to a drag force known as *dynamical friction*, which plays a major role in group and cluster dynamics. The timescale for dynamical friction can be written

$$\tau_{\mathrm{df}} \equiv \left( \frac{1}{v_\parallel} \frac{dv_\parallel}{dt} \right)^{-1} = \frac{v^3}{G^2 (m + m_f) \rho f(v/\sigma_v) \ln \Lambda} \ ,$$

where $\rho$ is the local mass density of field particles, $f$ is another function of order unity, and $\ln \Lambda$ is again the Coulomb logarithm. In principle, one could also compute a collective frictional timescale in a manner analogous to the collective relaxation timescale (Gurzadyan 1993, private communication). Maoz (1993) has recently computed the orbital energy dissipation from dynamical friction in inhomogeneous media, but his methodology does not return the actual force, which in general is not opposite to the motion of the test particle.

Perhaps more physical is the timescale for *orbital decay* defined as

$$\tau_{\mathrm{od}} \equiv \left(\frac{1}{R}\frac{dR}{dt}\right)^{-1} = \left(\frac{RdE/dR}{mv^2}\right)\tau_{\mathrm{df}} = \frac{3}{2}\left(\frac{\rho}{\bar{\rho}}+\frac{1}{3}\right)\tau_{\mathrm{df}} .$$

Unfortunately, this timescale does not always provide correct answers: 1) No orbital decay is predicted in zero density environments, whereas a satellite galaxy sitting just outside its parent galaxy will see its orbit decay, because of resonances with its parent (Lin & Tremaine 1983); 2) Although orbital decay should be slowed by tidal effects that reduce the test particle's mass, the contrary may occur with a satellite galaxy circling its parent, as the tides from the latter remove stars from the former, and these carry off energy and angular momentum, thus accelerating the orbital decay (Prugniel & Combes 1992). 3) If one throws a satellite right through a parent galaxy, the resultant energy loss by dynamical friction requires an unusually high Coulomb logarithm (10 or so) to match the results from PM simulations (Seguin, in these proceedings). In any event, the timescale for orbital decay in rich clusters is greater than a Hubble time for galaxies with $m < 10^{12} M_\odot$ (see Mamon 1985, §III), but starts to become important for groups of galaxies falling into these clusters.

Another outgrowth of dynamical friction is *orbital circularization*, whose timescale can be defined as

$$\tau_{\mathrm{oc}} = \left(\frac{1}{J_{\mathrm{circ}}(E)}\frac{dJ}{dt}\right)^{-1} ,$$

which Merritt (1985) finds to be shorter than the orbital decay time outside of the core radius of a cluster.

Tidal forces act on particles in a system *relative* to the full system itself. As such there are two types of tides acting on galaxies in groups and clusters: those caused by close encounters with other galaxies and those caused by variations in the gradient of the global group/cluster potential. The first type of tides (*collisional stripping*) has a timescale

$$\tau_{\mathrm{cs}} \equiv \left(\frac{1}{m}\frac{dm}{dt}\right)^{-1} = \langle(\Delta m/m)n\langle\sigma v\rangle\rangle^{-1} = \frac{\mathrm{Cst}}{nr_g^2 v_g} ,$$

where $\sigma$ is the collisional stripping cross-section, and the outer stars are assumed to follow elongated orbits (Richstone 1975; Dekel, Lecar & Shaham 1980).

Global potential tides depend strongly on the galaxy's orbit around the cluster. If the galaxy is phase locked in a nearly circular orbit around the cluster, it will feel a roughly constant tidal shear, and its tidal radius will be obtained by equating the tidal shear at a given radius in the galaxy with the gravitational pull that the full galaxy exerts on a star at that radius, plus an inertial term:

$$\Delta\left(\frac{GM(r)}{R^2}\right) = -\frac{Gm(r)}{r^2} + \Omega^2 r , \qquad (1)$$

yielding for $r \ll R$

$$\bar{\rho}_g(r_t) = \bar{\rho}_{\mathrm{cl}}(R)\left[2 - 3\frac{\rho_{\mathrm{cl}}(R)}{\bar{\rho}_{\mathrm{cl}}(R)} + \frac{V_p^2(R)}{V_{\mathrm{circ}}^2(R_p)}\right] , \qquad (2)$$

*i.e.*, the galaxy is tidally truncated at a radius $r_t$ where its mean density is of the order of the mean cluster density within the radius $R_p$ of closest approach of the galaxy (where $V_p$ and $V_{\rm circ}$ are the pericentric and circular velocities, respectively). Merritt (1984) has argued that central cD galaxies could not have spiralled in from outside the cluster cores, for otherwise these global potential tides would have seriously limited their sizes.

If the orbits are elongated, the instantaneous tide obtained from equation (1) is short lived and the galaxy experiences a *tidal shock* (Ostriker, Spitzer & Chevalier 1972). Using the impulse approximation (Spitzer 1958), in which the perturber moves with a constant relative velocity **V**, one can show (Mamon 1992a) that again for $r \ll R$

$$\bar\rho_g(r_t) = {\rm Cst}\, \bar\rho_{\rm cl}(R_p) f(\epsilon)\ ,$$

where $R_p$ is the pericentric of the galaxy's orbit, and $f(\epsilon)$ is a function of order unity of the galaxy's orbital eccentricity. This criterion is similar to that for circular orbits, but the constants are higher, because at given pericenter, a galaxy in a circular orbit must feel a more effective tide, since it is long-lived (Mamon 1987). Numerical simulations by Allen & Richstone (1988) confirm this result although other simulations by Merritt & White (1987) suggest that the tide is most efficient for some intermediate elongation at given pericenter, when this is within the nearly homogeneous region of the cluster. Note that the timescales for global potential tides are basically the orbital timescales divided by the typical mass-loss per passage through the cluster core.

The effectiveness of a tide is related to the maximum strength of the tide times the duration of this maximum tide. So, from equation (2) one gets

$$\Delta v \sim F_{\rm tid}\Delta t \sim \bar\rho_g \Delta t \sim \frac{2 - 3\rho_{\rm cl}/\bar\rho_{\rm cl} + V_p^2/V_{\rm circ}^2}{V_p/V_{\rm circ}}$$
$$\sim 3\left(1 - \frac{\rho_{\rm cl}}{\bar\rho_{\rm cl}}\right) - \left(1 - 3\frac{\rho_{\rm cl}}{\bar\rho_{\rm cl}}\right)\left(\frac{V_p}{V_{\rm circ}} - 1\right) \qquad {\rm for\ } V_p \gtrsim V_{\rm circ}\ .$$

Hence, the results of Merritt & White are understood, since when the cluster region is nearly homogeneous, the effective tide increases with increasing pericenter velocity, but not when the cluster density profile decreases sharply as outside the core of the Modified Hubble model used by Merritt & White.

The timescale for merging may be estimated from a merging cross-section, again as

$$\tau_m = n \left\langle \sigma v \right\rangle^{-1}\ .$$

Using Roos & Norman's (1979) numerically experimental cross-section, the merger time can be written (adapted from Mamon 1992a)

$$\tau_m = {\rm Cst}\left[nr_g^2 v_g K(v_{\rm cl}/v_g)\right]^{-1}\ , \tag{3}$$

where $n$ is the number density of galaxies, $r_g$ and $v_g$ are the galaxy half-mass radius and internal velocity dispersion, respectively, and where the merging efficiency $K$ is optimum for groups ($v_{\rm cl} \simeq v_g$), while for clusters it falls off as $v_{\rm cl}^3$. In groups as dense as Hickson's (1982) compact groups appear to be, merging ought to be extremely efficient, and the relatively low fraction of ellipticals indicates that chance alignments are contaminating the Hickson compact group catalog (Mamon 1992a). Despite their high velocity dispersions, rich clusters seem to be able to produce the right amount of mergers to produce elliptical morphologies, and moreover, merging is able to account for the morphology-density (Postman & Geller 1984) and morphology-radius (Whitmore & Gilmore 1991) relations (Mamon 1992a, and in these proceedings).

The physical processes described above compete in the evolution of the galaxy system. For example, merging leads to increased merger cross-sections, hence to a merging instability (Ostriker & Hausman 1977; see also the simulations of Carnevali, Cavaliere & Santangelo 1981 and the analytical formulation by Cavaliere, Colafrancesco & Menci 1992). However, this instability is slowed down by tidal processes which are usually thought to truncate galaxies of their outlying particles which become unbound (Mamon 1987). Yet, if the merging cross-section is related to galaxy half-mass radius (Aarseth & Fall 1980), and since the tidal processes for galaxies on elongated orbits or from collisions pump energy into the system, then the half-mass radius of those particles that remain bound to the galaxy should increase. The question remains whether the new half-mass radius is then greater or smaller than the old value, but this reviewer is not aware of any numerical study that has addressed this question yet.

In any event, it becomes necessary to run numerical simulations to see how groups and clusters evolve. The reader is referred to Athanassoula, Friedli, and Scholl (all three in these proceedings) for presentations of the numerical techniques, and to Mamon (1990) for comparison of the results on merging in dense groups from different techniques.

The principal results are as follows: The dynamics of clusters is now understood to depend strongly on the primordial density fluctuation spectrum (West, Oemler & Dekel 1988). Galaxies *overmerge* in clusters and possibly in dense groups, when simulated with collisionless particles (*e.g.,* White *et al.* 1987) and this overmerging seems to be caused by the fact that the particles in the halos of galaxies relax rapidly with the intergalactic particles within the core of the system (Villumsen 1993). This is not seen in simulations where gas is included (Evrard, Summers & Davis 1992; Katz & White 1993), presumably because the gas sinks to the bottom of the halo potential wells and deepens these wells, which thus avoid merging with one another. Dense groups of galaxies witness rapid merging and coalesce into a single elliptical galaxy (Carnevali *et al.* 1981; Barnes 1985; Mamon 1987; Barnes 1989; Lima Neto, in these proceedings).

A detailed comparison of the results on groups (Mamon 1990) showed that the different numerical studies of groups produced comparable rates of merging. In an interesting study, Garcia & Athanassoula (in these proceedings) have gone one step further by simulating the same groups by the various methods (*explicit-physics* with one particle per galaxy and the physics of interactions [§2] explicitly included, and the *self-consistent* methods in which galaxies are constituted of many particles). They point out a discrepancy between the merging cross-section of Roos & Norman (1979) which seems too high, whereas that of Aarseth & Fall (1980), curiously derived from the former, seems to give decent results.

Whereas simulations by Cavaliere *et al.* (1982), Barnes (1985) and Mamon (1987) all show that dense groups survive longer if the dark matter is distributed in a common envelope, the contrary as been found in recent simulations by Athanassoula & Makino (1993). What causes this discrepancy? If galaxies have individual halos, merging is direct (Mamon 1987), and the merging rate is proportional to the merger cross-section, and hence to the square of the galaxy half-mass radius (eq. [3]). Usually, the individual halos of dark matter are more extended by construction than the luminous matter, and the merging cross-section is increased by a factor of nearly 100, more than compensating the positive effects of dynamical friction on the merging rate when the dark matter is in a common envelope (Mamon 1987). But in Athanassoula & Makino's simulated galaxies, the dark matter halos have the same matter distribution as the luminous matter (simply scaled up), hence the presence of dark matter halos did not increase the merger cross-sections, while the runs with a common envelope merged faster thanks to dynamical friction.

## 3. Substructure in Clusters

Although perhaps 30% of clusters exhibit *large-scale* substructure (*e.g.*, Jones & Forman 1992), various statistical studies on optical data (Salvador-Solé, Sanromà & González-Casado 1993; Salvador-Solé, González-Casado & Solanes 1993) and recent *ROSAT* observations (*e.g.*, White, Briel & Henry 1993) show that *small-scale* substructure is present in a majority of clusters. Three recent studies (Richstone, Loeb & Turner 1992; Lacey & Cole 1993; Kauffmann & White 1993) have attempted to obtain constraints on the density parameter $\Omega$ from the frequency of substructure in clusters. The idea is that if $\Omega < 1$, then structures in the Universe collapse from their initial Hubble expansion at epochs $z \simeq 1/\Omega$, while if $\Omega = 1$, structures keep collapsing today (Gott & Rees 1975; Richstone, Loeb & Turner 1992). The first two of the three studies conclude to $\Omega_0 > 0.5$, while as noted in the third (Kauffmann & White), the problem is that the dynamical survival time of substructures is only guessed (Richstone *et al.*) or treated too simplistically (Lacey & Cole).

In fact, one can do better, and consider as two extreme cases the accretion of a group into a cluster, and the merging of two similar-mass clusters with the decoupling of their dense cores. One can then compare the ability of these two extreme scenarios to produce small-scale substructure of a mass-fraction of say 5 or 10%. Preliminary calculations indicate that groups are destroyed by tides from the global cluster potential in one passage through the cluster core, whereas the stripped cores of clusters are able to survive such tides for a few orbital periods. The difference arises simply because groups have lower mean density than the detached cores of clusters, and thus are easier to destroy (see González-Casado, Mamon & Salvador-Solé 1993). Moreover, while the more massive substructures survive tides better at first passage through the cluster core, their orbits decay faster by dynamical friction, thus reducing their lifetime $\Delta t$. One thus expects a small range of mass fractions, which is consistent with the observations (González-Casado, Mamon & Salvador-Solé 1993).

## 4. X-ray Observations of Groups

Very recently, a diffuse hot intergalactic background has been discovered in two groups with pointed observations of the *ROSAT* satellite: the loose group NGC 2300 (Mulchaey *et* *al.* 1993) and the compact group HCG 62 (Ponman & Bertram 1993). In both groups, the diffuse IGM has a temperature of about 1 keV (to within 15%) although the compact group has a dip in its central temperature, probably caused by a cooling flow, since the cooling time in the center is found to be short. Also, both groups have low metallicity compared to clusters, consistent with nearly primordial gas, rather than enriched by supernova ejecta.

The NGC 2300 group is claimed to have a rather high dynamical mass within a radius of $165\,h^{-1}$ kpc (Mulchaey *et al.* 1993), and consequently a very low baryonic fraction, 4%, consistent with the constraints from big-bang nucleosynthesis. Note that the NGC 2300 group has very uncertain parameters: the X-ray surface brightness profile is so poorly constrained out to $R = 45'$ that its background-subtracted asymptotic slope is uncertain to at least a factor of three (Henriksen & Mamon 1993). This implies an uncertainty of a factor two in the total mass within a radius of $25'$, and the resultant gas fractions range between 14% and 24% (Henriksen & Mamon), thus higher than the limits obtained from nucleosynthesis. Moreover, for low asymptotic slope, the baryonic fraction increases with radius, and conversely for high slopes (Henriksen & Mamon 1993). Hence the need for more extended X-ray observations with, for example off-center pointings, which are indeed planned (Burstein 1993, private communication).

## 5. A Unified Scheme for Groups

Groups of galaxies have often been used to argue for low values of the cosmological density parameter $\Omega$, since their mass-to-light ratios are $\lesssim 10\%$ of the required value to close the Universe. However, these mass analyses assume that groups are virialized entities. It has been shown that groups are rarely virialized (Byrd & Valtonen 1985; Giuricin *et al.* 1988). Diaferio *et al.* (1993) go further and say that the observational properties of the groups that Ramella, Geller & Huchra (1989) extracted from the CfA slice are compatible with a single collapsing group observed from different viewing angles.

Simple cosmological theory provides more insight into the evolution of the observable properties of groups. A homogeneous isolated group should see its size evolve as shown in Figure 1a. It first follows the Hubble expansion, then decouples from this expansion and turns around, collapses and subsequently virializes. Applying the virial theorem, one derives a virial mass $M_V = R_V V_V^2/G$ and crossing time $t_V = R_V/V_V$ to within known constants of order unity. In an important paper, Giuricin *et al.* (1988) have shown how to compute the observable mass and crossing time of a group in terms of its cosmo-dynamical state. Figures 1b, 1c, and 1d show the biases in observable velocity dispersion, mass, and crossing time using their analysis to the idealized evolution depicted in Figure 1a. The dotted track is for groups made of point mass galaxies, while the solid track is for extended galaxies, which reach a terminal velocity at group collapse (because the smoothed potential is flat at the center), and after virialization, dissipate their orbital energy by dynamical friction against their common massive halo (merged from their individual halos after group collapse).

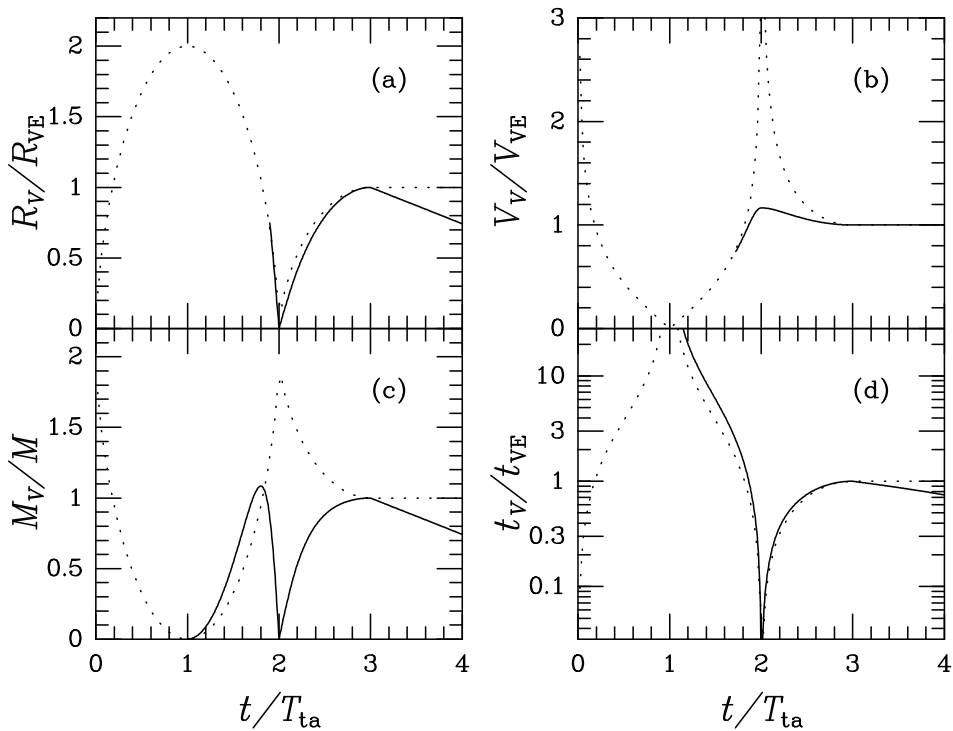

**Figure 1.** Time evolution of bias in observed virial radius (a), velocity dispersion (b), mass (c), and crossing time (d), relative to virial equilibrium (VE), where $T_{\rm ta}$ is the turnaround time. The *dotted curves* show the evolution for point-masses, while the *full curves* show the effects of softened potentials and orbital energy dissipation by dynamical friction (starting at $t = 3\,T_{\rm ta}$).

In Figure 2a is shown the theoretical evolution of a group in the space ($M_V/M$ versus $t_V/t_0$), which can be understood to be analogous to an evolutionary track in a Hertzprung-Russell diagram for stars. To compare with parameters from observed groups, we must link the groups to a same mass scale, and do so by assuming that the *true* $M/L$ is constant from group to group and independent of its cosmo-dynamical state. In Figures 2b, c, and d, we plot the observed group parameters ($M_V/L$ vs. $H_0 t_{\rm cr}$, for groups of different multiplicities, and superpose the theoretical evolutionary track, adjusting the $y$-axis with the high multiplicity groups of Figure 2b, while the $x$-axis scaling is imposed by theory. The groups are taken from the Gourgoulhon, Chamaraux & Fouqué (1992) catalog of groups, the largest available in the literature, but the results below have been checked with Tully's (1987) groups.

The high-multiplicity groups fit the theoretical tracks very well. A one proceeds to lower multiplicities, the statistical noise in the mass-to-light ratio and crossing time estimates increases, but so does the probability for chance alignments, which make the groups appear smaller while conserving on the average their velocity dispersion. Although precise assignments of group cosmo-dynamical states is difficult because of statistical noise, one can nevertheless get a handle on which groups are unbound (above theoretical track), which are still in their expansion phase (upper-right handle of track), which are near turnaround (lower-right handles of track), which are collapsing (central handle), which are near maximum collapse (first lower-left handle), and those that are virialized (second lower-left handle). The theoretical track thus represents a slice through the *fundamental surface* (which is curved) of groups, where the third axis is total group luminosity.

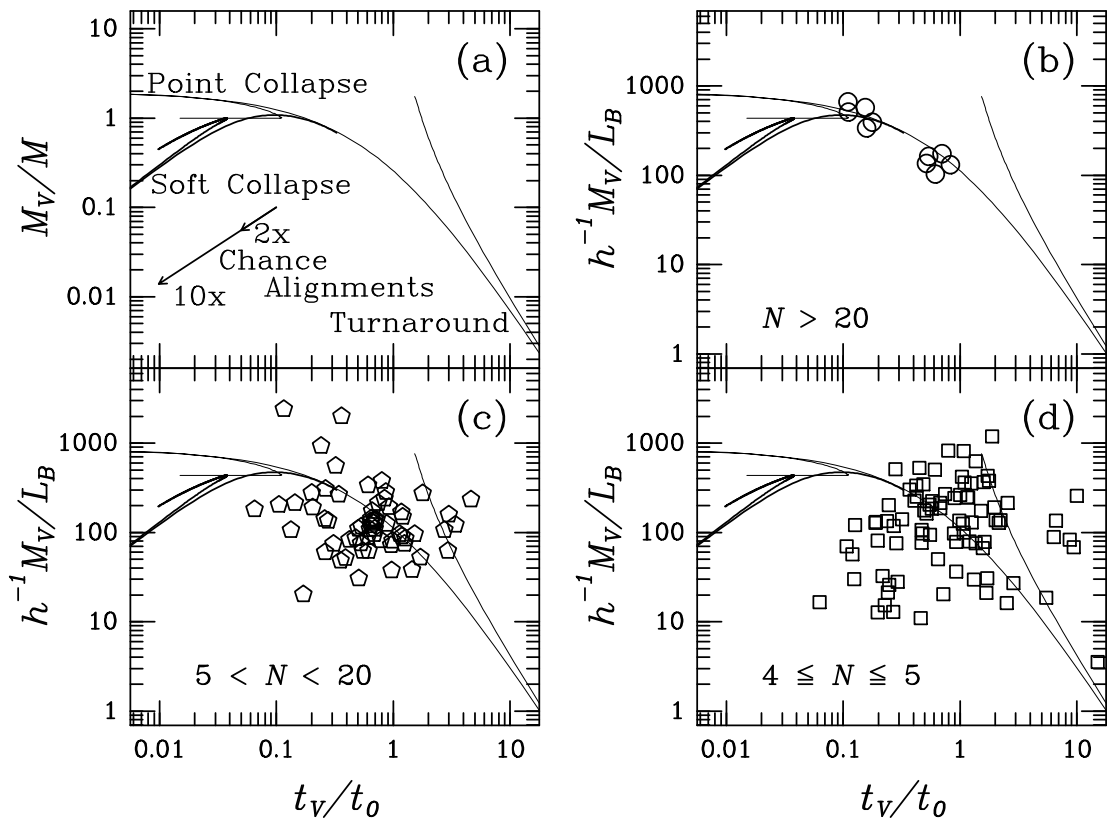

**Figure 2.** Mass, scaled to total mass (a) or total blue luminosity (b, c, and d), versus crossing time (in units of the age of the Universe for $\Omega = 1$, while for $\Omega = 0.2$ the points should be displaced to the left by 0.1 decade). The *polygons* (b, c, and d) represent the loose groups from Gourgoulhon, Chamaraux & Fouqué (1992). The *thin curves* are the theoretical point-mass evolutionary tracks, while the *thick curves* are the same for softened potentials and allowing for orbital energy dissipation after virialization. In (b, c, and d), these curves are scaled to mass-to-light ratios assuming that all groups have a true $M/L = 440\,h$.

The true $M/L$ is obtained by extrapolating to the early virialized state (before dissipation of orbital energy, which occurs at nearly constant velocity dispersion since the common halo should have near constant circular velocity). The Gourgoulhon et al. groups then have $M_{\rm true}/L = 440\,h$, much higher than the median $M/L = 130\,h$, for the groups of $N \geq 4$ members (the mass estimate used here is the median of the non-weighted virial, weighted virial, and projected masses). In other words, *the mass-to-light ratios of groups are severely underestimated because most groups are still relatively near their turnaround phase*. This points to $\Omega \simeq 0.3$ obtained by extrapolating Loveday et al.'s (1992) galaxy luminosity function to $(M/L)_{\rm closure} = 1560\,h$. Barnes (1985) showed similar plots as in Figure 2 for simulated groups of 5 galaxies starting from turnaround and also concluded for mass estimates of observed groups too low by a factor three or more, but attributes this to mass segregation between galaxies and dark matter at group collapse instead of the bias near turnaround advocated here.

In any event, *no groups in the loose group catalog has yet completed its collapse*, not even the Virgo cluster included in the catalog, whose outer members are still collapsing onto the virialized core. Although this conclusion is in accord with the single collapsing state advocated by Diaferio *et al.* (1993), the present analysis allows a range in cosmo-dynamical states. Now, if $\Omega = 0.1$, then structures would form at $z \simeq 10$, and there should be few collapsing groups today (Gott & Rees 1975). *The fact that all groups are in a young cosmo-dynamical state, thus points to a high $\Omega$, perhaps close to unity.* Details of this analysis will be found in Mamon (1993b).

## 6. Real vs. Accidental Compact Groups

The nature of the compact groups such as those cataloged by Hickson (1982) has been a matter of much debate. On one hand, the high level of galaxy-galaxy interaction is becoming increasingly evident as the numerous observational studies of compact groups progress. To summarize briefly, compact group galaxies are often morphologically (Hickson 1990; Mendes de Oliveira 1992) or kinematically (Rubin, Hunter & Ford 1991) disturbed.

However, various theoretical and statistical arguments point against the 3D high density of *the majority* of Hickson's compact groups (once the obvious interlopers with discordant redshifts are culled out). Indeed, 1) It is hard to understand how bound dense groups form in sufficient numbers, given their short survival times against depletion from galaxy mergers (see Mamon 1987 for a statistical appraisal of the survival of dense groups against mergers). 2) Simulations of virialized dense groups (Mamon 1987) show rapid evolution of the bright-end of the luminosity function, in sharp contrast with what is observed for the ensemble of Hickson groups (Mamon 1986). This argument implies that most compact groups could not have been dense in 3D for over 1 or 2 Gyr.

The alternative to compact groups that are dense in 3D are compact groups caused by chance alignments of loose group galaxies along the line of sight. Simulations of *virialized* loose groups have shown that such 1D chance alignments are roughly 10 times more frequent than the formation of 3D dense groups by 2-body processes (Mamon 1992b). Moreover, these chance alignments are binary-rich as only one-quarter is composed of 4 or more unrelated galaxies (Mamon 1992b). A rule of 3 on the frequency of binaries in chance alignments, shows that the fraction of interacting galaxies in groups is consistent with the observed high frequency of 63% (Rubin *et al.* 1991) of compact group galaxies with abnormal internal kinematics, once one folds in a fraction of 10% of truly dense groups in Hickson's sample (Mamon 1992b).

A recent detailed morphological analysis of compact group galaxies (Mendes de Oliveira 1992) shows that 35% of Hickson's compact groups have 3 or more interacting galaxies, whereas the prediction from *only* chance alignments is 19% to 27% (Mamon 1993a). The discrepancy gets worse once the subsample of 16 compact groups with kinematical data is considered, as Mendes de Oliveira finds that 75% of these groups have 3 or more interacting galaxies, combining her morphological analysis with Rubin *et al.*'s (1991) kinematical analysis. But if one-third of the accordant-redshift Hickson compact groups are real while the remainder are binary-rich chance alignments, one then obtains 55% of Hickson's compact groups showing 3 or more interacting galaxies (assuming that dense triplets and quartets *always* show morphological or kinematical interactions). Considering that some of the interactions seen in the sample of 16 could be caused by accretion of dwarf galaxies rather than interaction between galaxies bright enough to be listed in Hickson's catalog, the discrepancy is not strong enough, in this reviewer's opinion, to rule out that the majority of compact groups are caused by chance alignments.

Where do compact groups lie in the $M_V/L$ vs. $H_0 t_{cr}$ diagram (see Mamon 1993b for details)? This is shown in Figure 3, for the accordant redshift compact groups of four or more members (Hickson et al. 1992), which lie within three regions: 1) The low-velocity dispersion compact groups (lower right) are mostly chance alignments within collapsing loose groups. 2) The intermediate velocity dispersion compact groups are mostly loose groups near full cosmological collapse. 3) The high velocity dispersion compact groups (upper left) are mostly virialized loose groups. The previously unexplained morphology-velocity dispersion relation in compact groups (Hickson, Kindl & Huchra 1988) is then attributable to the fact that only the high velocity dispersion compact groups have had enough time to reach virialization and hence witness rapid merging within them to form ellipticals.

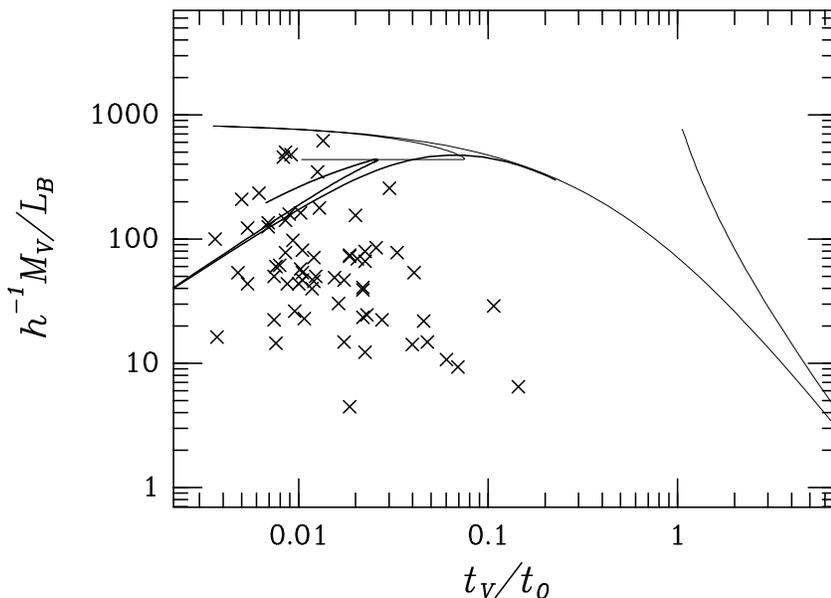

**Figure 3.** Mass-to-light ratio versus crossing time of compact groups (*crosses*) and theoretical evolutionary tracks (see fig. 2).

## 7. Real vs. Accidental Binaries

Binary galaxies have been often used to probe the existence and extent of galaxy halos (Turner 1976b; White et al. 1983; Schweizer 1987; Charlton & Salpeter 1991), with contradictory results. Indeed, if binary halos overlap, their global kinematics should be altered relative to non-overlapping halos, which orbit in the same manner as point masses. Because a substantial fraction ($\simeq 40\%$) of binaries reside within groups, it is important to assess what fraction of the binaries within groups are caused by chance alignments and which fraction are truly bound pairs. Brieu & Mamon (1993) have employed simulations of virialized groups looking for pairs in projection meeting the binary isolation criteria of Turner (1976a) or Schweizer (1987). They find that the fraction of accidental binaries is between 40% and 80%, depending on whether galaxies have individual halos or not. Moreover, the correlation of the observational quantities that are the physical projected separation and the radial velocity difference are similar between real and accidental pairs. There is no way to distinguish between the two sets by selecting binaries with small projected separation or small radial velocity difference. Thus, *the isolation criteria used to select binaries are insufficient to select real binaries*, and one has to completely cull out the binaries within group to avoid being swamped by accidental pairs. The weak point of this analysis is that it is based upon virialized groups, whereas loose groups are not virialized (see §5), which should in principle alter the internal kinematics of groups and hence of the binaries appearing in projection.